\documentclass[11pt,conference,twocolumn]{IEEEtran}
\usepackage{ieeefig}
\usepackage{amssymb}
\usepackage{graphicx}
\usepackage{amsmath}
\usepackage{epsfig}
\usepackage{color}
\usepackage{url}

\newtheorem{lemma}{Lemma}
\newtheorem{theorem}{Theorem}

\begin{document}

\title{Efficient Parallel and Out of Core Algorithms for Constructing Large Bi-directed de Bruijn Graphs}

\author{
\IEEEauthorblockN{Vamsi Kundeti and Sanguthevar Rajasekaran and Hieu Dinh}
\IEEEauthorblockA{Department of Computer Science and Engineering\\
University of Connecticut, Storrs, CT 06269\\
emails: \{vamsik, rajasek, hieu\}@engr.uconn.edu}
\IEEEauthorblockN{\and Matthew Vaughn and Vishal Thapar}
\IEEEauthorblockA{Cold Spring Harbor Laboratory \\
Cold Spring, NY\\
emails:\{mvaughn, vthapar\}@cshl.edu;}
}




\maketitle               

\begin{abstract}
Assembling genomic sequences from a set of overlapping reads is one of the
most fundamental problems in computational biology. Algorithms addressing the
assembly problem fall into two broad categories -- based on the data structures
which they employ. The first class uses an overlap/string graph and the second 
type uses a de Bruijn graph. However with the recent advances in short read 
sequencing technology, de Bruijn graph based algorithms seem to play a vital role in 
practice. 

Efficient algorithms for building these massive de Bruijn graphs are very essential 
in large sequencing projects based on short reads. In~\cite{par_bidirected_graph}, an $O(n/p)$ time parallel algorithm has been given for this problem. Here $n$ is the size of the input and $p$ is the number of processors. This algorithm enumerates all possible bi-directed
edges which can overlap with a node and ends up generating $\Theta(n\Sigma)$ messages. 

In this paper we  present a $\Theta(n/p)$ time parallel algorithm with a 
communication complexity equal to that of parallel sorting and is not sensitive to $\Sigma$. 
The generality of our algorithm makes it very easy to extend it even to the 
out-of-core model and in this case it has an optimal I/O complexity of $\Theta(\frac{n\log(n/B)}{B\log(M/B)})$. 
We demonstrate the scalability of our parallel algorithm on a SGI/Altix computer. A comparison of our algorithm with that of~\cite{par_bidirected_graph} reveals that our algorithm is faster. We also provide efficient algorithms for the
bi-directed chain compaction problem.
\end{abstract}


\section{Introduction}

\PARstart The {\em genomic sequence} of an organism is a string from the alphabet $\Sigma=\{A,T,G,C\}$.
This string is also referred as the Deoxyribonucleic acid (DNA) sequence. DNA sequences exist as 
complementary pairs ($A-T$, $G-C$) due to the double strandedness of the underlying DNA structure. 
Several characteristics of an organism are encoded in its DNA sequence, thereby reducing the 
biological analysis of the organism to the analysis of its DNA sequence. Identifying the unknown 
DNA sequence of an organism is known as {\em de novo sequencing} and is of fundamental biological 
importance. On the other hand the existing sequencing technology is not mature enough to identify/read the 
entire sequence of the genome -- especially for complex organisms like the mammals. However small 
fragments of the genome can be read with acceptable accuracy.  The {\em shotgun} method employed 
in many sequencing projects breaks the genome randomly at several places and generates several
small fragments ({\em reads}) of the genome. The problem of reassembling all the fragmented reads into 
a small sequence close to the original sequence is known as the {\em Sequence Assembly }(SA) problem. 

Although the SA problem seems similar to the {\em Shortest Common Super string} (SCS) problem, there are
in fact some fundamental differences. Firstly, the genome sequence might contain several repeating
regions. However, in any optimal solution to the SCS problem we will not be able to find repeating 
regions -- because we want to minimize the length of the solution string. In addition to the repeats, there 
are other issues such as errors in reads and double strandedness of the reads which make the 
reduction to SCS problem very complex. 

The literature on algorithms to address the SA problem can be classified into two broad categories. 
The first class of algorithms model a read as a vertex in a directed graph -- known as the 
{\em overlap graph}~\cite{myers95}. The second class 
of algorithms model every substring of length $k$ (i.e., a $k$-mer) in a read as a vertex in a (subgraph of)
a {\em de Bruijn} graph~\cite{pevzner01}.

In an overlap graph, for every pair of overlapping reads, directed edges are 
introduced consistent with the orientation of the overlap. Since the transitive edges in the 
overlap graph are redundant for the assembly process they are removed and the resultant graph 
is called the {\em string graph}~\cite{myers95}. The edges
of the string graph are classified into {\em optional}, {\em required} and {\em exact}. The 
SA problem is reduced to the identification of a shortest walk in the string graph which includes all the 
required and exact constraints on the edges. Identifying such a walk -- {\em minimum $S$-walk}
-- on the string graph is known to be \textsf{NP}-hard~\cite{bidirected_graph}. 

When a de Bruijn graph is employed, we model every substring of length $k$ (i.e., a $k$-mer) in a read as a vertex ~\cite{pevzner01}.
A directed edge is introduced between two $k$-mers 
if there exists some read in which these two $k$-mers overlap by exactly $k-1$ symbols. Thus
every read in the input is mapped to some path in the de Bruijn graph. The SA problem
is reduced to a {\em Chinese Postman Problem} (CPP) on the de Bruijn graph, subject to the 
constraint that the resultant CPP tour include all the paths corresponding to the reads.
This problem is also known to be \textsf{NP}-hard. Thus solving the SA problem exactly on 
both these graph models is intractable. 

Overlap graph based algorithms were found to perform better 
(see ~\cite{PCAP}~\cite{CELERA}~\cite{ARCHANE}~\cite{PHRAP}) with Sanger based read methods. Sanger
methods produce reads typically around $1000$ base pairs long. However these can produce significant
read errors. To overcome the issues with Sanger reads new read technologies such as the 
pyrosequencing (454sequencing) have emerged. These read technologies can produce reliable and 
accurate genome fragments which are very short (up to $100$ base-pairs long). On the other hand short read
technologies can increase the number of reads in the SA problem by a large magnitude. Overlap based
graph algorithms do not scale well in practice since they represent every read as a vertex. De Bruijn
graph based algorithms seem to handle short reads very efficiently (see~\cite{velvet08}) in practice
compared to the overlap graph based algorithms. However the existing sequential algorithms~\cite{velvet08} 
to construct these graphs are sub-optimal and require significant amounts of memory. This limits the
applicability of these methods to large scale SA problems. In this paper we address this issue and present
algorithms to construct large de Bruijn graphs very efficiently. Our algorithm is optimal in the sequential,
parallel and out-of-core models.  A recent work by Jackson and Aluru~\cite{par_bidirected_graph} yielded
parallel algorithms to build these de Bruijn graphs efficiently. They present a parallel algorithm 
that runs in $O(n/p)$ time using $p$ processors (assuming that $n$ is a constant-degree ploynomial in $p$).
The {\em message complexity} of their algorithm is $\Theta(n\Sigma)$. By message complexity we mean
the total number of messages (i.e., $k$-mers) communicated by all the processors in the entire algorithm.
One of the major contributions of our
work is to show that we can accomplish this in $\Theta(n/p)$ time with a message complexity of $\Theta(n)$. An experimental
comparison of these two algorithms on an SGI Altix machine shows that our algorithm is considerably faster.
In addition, our algorithm works optimally in an out-of-core setting. In particular, our algorithm requires only
$\Theta(\frac{n\log(n/B)}{B\log(M/B)})$ I/O operations.

The organization of the paper is as follows. In Section~\ref{sec:prelim} we introduce some preliminaries
and define a bi-directed de Bruijn graph formally. Section~\ref{sec:our_algo} discusses our main algorithm
in a sequential setting. Section~\ref{sec:parallel} and Section~\ref{sec:out-of-core} show how our main 
idea can easily be extended to parallel and out-of-core models optimally. In Section~\ref{sec:remarks} 
we provide some remarks on the parallel algorithm of Jackson and Aluru~\cite{par_bidirected_graph}.
Section~\ref{sec:simplification} gives algorithms to perform the {\em simplification}
operation on the bi-directed de Bruijn graph. Section~\ref{sec:velvet} discusses how our {\em simplified}
bi-directed de Bruijn graph algorithm can replace the graph construction algorithm in a popular sequence
assembly program VELVET~\cite{velvet08}. Finally we present experimental results in Section~\ref{sec:exp_results}.

\section{Preliminaries}
\label{sec:prelim}
Let $s \in \Sigma^{n}$ be a string of length $n$. Any substring $s_j$ (i.e., $s[j]s[j+1]\ldots s[j+k-1], n-k+1\geq j\geq 1$) of 
length $k$ is called a $k-$mer of $s$. The set of all $k-$mers of a given string $s$ is called the $k-$spectrum of 
$s$ and is denoted by $\mathbb{S}(s,k)$. Given a $k-$mer $s_j$, $\bar{s_j}$ denotes the {\em reverse complement} of 
$s_j$ (e.g., if $s_j = AAGTA$ then $\bar{s_j} = TACTT$). Let $\leq$ be the partial ordering among the strings of equal 
length, then $s_i \leq s_j$ indicates that the string $s_i$ is lexicographically smaller than $s_j$. Given any $k-$mer $s_i$, 
let $\hat{s_i}$ be the lexicographically smaller string between $s_i$ and $\bar{s_i}$. We call 
$\hat{s_i}$ the {\em canonical} $k-$mer of $s_i$. In other words, if $s_i \leq \bar{s_i}$ then $\hat{s_i} = s_i$ 
otherwise $\hat{s_i} = \bar{s_i}$. A $k-$molecule of a given $k-$mer $s_i$ is a tuple $(\hat{s_i},\bar{\hat{s_i}})$ 
consisting of the canonical $k-$mer of $s_i$ and the reverse complement of the canonical $k-$mer. In the rest of
this paper we use the terms positive strand and canonical $k-$mer interchangeably. Likewise the non-canonical
$k-$mer is referred to as the negative strand.

\begin{figure}
\begin{center}
\includegraphics[scale=0.6]{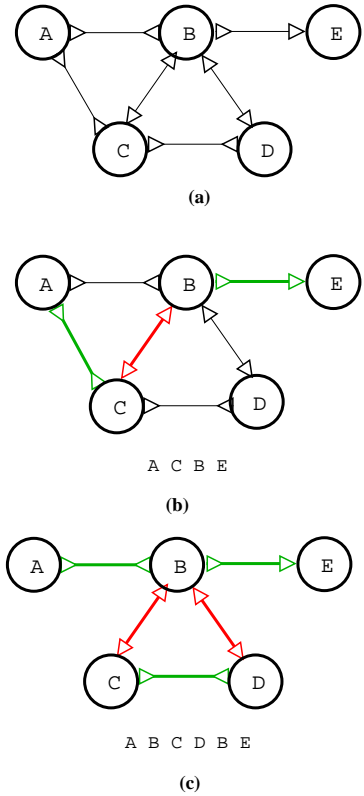}
\end{center}
\caption{Bi-directed graph and bi-directed walks}
\label{fig:bi-walk-example}
\end{figure}

A {\em bi-directed} graph is a generalized version of a standard directed graph. In a directed graph every 
edge has only one arrow head ($\text{--}\rhd$ or $\lhd\text{--}$). On the other hand in a bi-directed graph 
every edge has two arrow heads attached to it ($\lhd\text{--}\rhd$, $\lhd\text{--}\lhd$,$\rhd\text{--}\lhd$ or $\rhd\text{--}\rhd$).
Let $V$ be the set of vertices and 
$E = \{(v_i,v_j,o_1,o_2) | v_i,v_j\in V \wedge o_1,o_2\in\{\lhd,\rhd\}\}$ be the set of bi-directed edges 
in a bi-directed graph $G(V,E)$. For any edge $e = (v_i,v_u,o_1,o_2)\in E$,
$o_1=e[o^+] $ and $o_2=e[o^-]$ refer to the orientations of the arrow heads on the vertices $v_i$ and $v_j$,
respectively. A {\em walk} $W(v_i,v_j)$ between two 
nodes $v_i,v_j \in V$ of a bi-directed graph $G(V,E)$ is a sequence 
$v_i,e_{i_1},v_{i_1},e_{i_2},v_{i_2},\ldots, v_{i_m},e_{i_{m+1}},v_j$, such that for every intermediate 
vertex $v_{i_l},1\leq l \leq m$ the orientation of the arrow head on the incoming
edge adjacent on $v_{i_l}$ is opposite to the orientation of the arrow head on the out going edge.
To make this clearer, let $e_{i_l},v_{i_l},e_{i_{l+1}}$ be a sub-sequence in the walk $W(v_i,v_j)$.
If $e_{i_l}=(v_{i_{l-1}},v_{i_l},o_1,o_2), e_{i_{l+1}} = (v_{i_{l}},v_{i_{l+1}},o_1',o_2')$
then for the walk to be valid it should be the case that $o_2 \neq o_1'$. Figure~\ref{fig:bi-walk-example}$(a)$ 
illustrates an example of a bi-directed graph. Figure~\ref{fig:bi-walk-example}$(b)$ shows a simple
bi-directed walk between the nodes $A$ and $E$. Bi-directed walk between two nodes may not be simple.
Figure~\ref{fig:bi-walk-example}$(c)$ shows a bi-directed walk between $A$ and $E$ which is not 
simple -- because $B$ repeats twice.

A de Bruijn graph $D^k(s)$ of order $k$ on a given string $s$ is defined as follows. 
The vertex set $V$ of $D^k(s)$ is defined as the $k-$spectrum of $s$ (i.e. $V = \mathbb{S}(s,k)$). 
We use the notation $suf(v_i,l)$ ($pre(v_i,l)$, respectively) to denote the suffix (prefix, respectively) of length $l$ in the string $v_i$.
Let the symbol $\circ$ denote the concatenation operation between two strings. The set of directed edges $E$ of $D^k(s)$ 
is defined as follows: $E = \{(v_i,v_j) | suf(v_i,k-1)=pre(v_j,k-1) \wedge v_i[1]\circ suf(v_i,k-1)\circ v_j[k] 
\in \mathbb{S}(s,k+1) \}$. We can also define de Bruijn graphs for sets of strings as follows. If $S=\{s_1,s_2\ldots s_n\}$ is any set
of strings, a de Bruijn graph $B^k(S)$ of order $k$ on $S$ has $V = \displaystyle\cup_{i=1}^{n} \mathbb{S}(s_i,k)$
and $E = \{(v_i,v_j) | suf(v_i,k-1) = pre(v_j,k-1) \wedge \exists \, l : v_i[1]\circ suf(v_i,k-1)\circ v_j[k]\in \mathbb{S}(s_l,k+1)\}$.
To model the double strandedness of the DNA molecules we should also consider the reverse 
complements ($\bar{S} =\{\bar{s_1},\bar{s_2}\ldots \bar{s_n}\}$) while we build the de Bruijn graph. 

To address this a bi-directed de Bruijn graph $BD^k(S\cup \bar{S})$ has been suggested in ~\cite{bidirected_graph}. 
The set of vertices $V$ of $BD^k(S\cup \bar{S})$ consists of all possible $k-$molecules from $S\cup\bar{S}$. 
The set of bi-directed edges for $BD^k(S\cup \bar{S})$ is defined as follows. Let $x,y$ be two $k-$mers 
which are next to each other in some input string $z \in S\cup \bar{S}$. Then an edge is introduced between the 
$k-$molecules $v_i$ and $v_j$ corresponding to $x$ and $y$, respectively. Please note that two consecutive $k-$mers in some input 
string always overlap by $k-1$ symbols. The converse need not be true. The orientations of the arrow heads on the 
edges are chosen as follows. If both $x$ and $y$ are the positive strands in $v_i$ and $v_j$, respectively,
then an edge $(v_i,v_j,\rhd,\rhd)$ 
is introduced. If $x$ is the positive strand in $v_i$ and $y$ is the negative strand in $v_j$ an edge $(v_i,v_j,\rhd,\lhd)$ 
is introduced. Finally, if $x$ is the negative strand in $v_i$ and $y$ is the positive strand in $v_j$ an 
edge $(v_i,v_j,\lhd,\rhd)$ is introduced. 

\begin{figure}
\begin{center}
\includegraphics[scale=0.6]{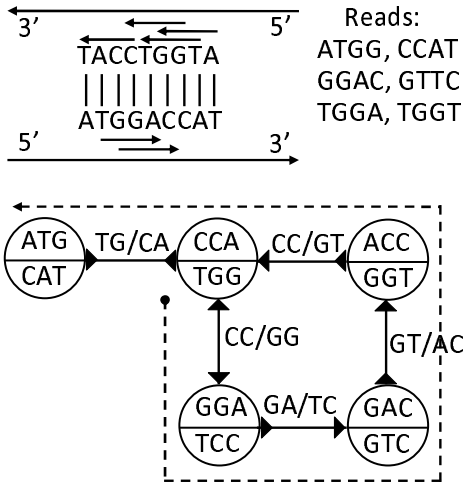}
\end{center}
\caption{Bi-directed de Bruijn graph example}
\label{fig:bi-de-graph-hieu}
\end{figure}

Figure~\ref{fig:bi-de-graph-hieu} illustrates a simple example of
the bi-directed de Bruijn graph of order $k=3$ from a set of reads
$ATGG,CCAT, GGAC, GTTC, TGGA$ and $TGGT$ observed from a DNA
sequence $ATGGACCAT$ and its reverse complement $ATGGTCCAT$.
Consider two $3-$molecules $v_1=(GGA,TCC)$ and $v_2=(GAC,GTC)$.
Because the positive strand $x=GGA$ in $v_1$ overlaps the positive
strand $y=GAC$ in $v_2$ by string $GA$, an edge
$(v_1,v_2,\rhd,\rhd)$ is introduced. Note that the negative strand
$GTC$ in $v_2$ also overlaps the negative strand $TCC$ in $v_2$ by
string $TC$, so the two overlapping strings $GA$ and $TC$ are drawn
above the edge $(v_1,v_2,\rhd,\rhd)$ in
Figure~\ref{fig:bi-de-graph-hieu}. A bi-directed walk on the example
bi-directed de Bruijn graph as illustrated by the dash line is
corresponding to the original DNA sequence with the first letter
omitted $TGGACCAT$. We would like to remark that the parameter $k$
is always chosen to be odd to ensure that the forward and reverse
complements of a $k$-mer are not the same.

\section{Our algorithm to construct bi-directed de Bruijn graphs}
\label{sec:our_algo}
In this section we describe our algorithm {\sf BiConstruct} to construct a bi-directed de Bruijn 
graph on a given set of reads. The following are the main steps in our algorithm to build the 
bi-directed de Bruijn graph. Let $R_f=\{r_1,r_2\ldots r_n\}, r_i\in \Sigma^r$ be the input set of reads. 
$\bar{R_f} = \{\bar{r_1},\bar{r_2}\ldots \bar{r_n}\}$ is a set of reverse complements. 
 Let $R^{*} = R_f \cup \bar{R_f}$ and $R^{k+1} = \displaystyle\cup_{r\in R^{*}}\mathbb{S}{(r,k+1)}$. 
$R^{k+1}$ is the set of all $(k+1)$-mers from the input reads and their reverse complements.
\begin{itemize}
\item {\sf [STEP-1] }{\bf Generate canonical edges:} 
Let $(x,y) = (z[1\ldots k], z[2\ldots k+1])$ be the $k-$mers corresponding to a 
$(k+1)$-mer $z[1\ldots k+1]\in R^{k+1}$. Recall that $\hat{x}$ and $\hat{y}$ are the canonical
$k-$mers of $x$ and $y$, respectively. Create a canonical bi-directed edge $(\hat{v_i}, \hat{v_j}, o_1, o_2)$ 
for each $(k+1)$-mer as follows.
$$
\begin{array}{l}
(\hat{v_i}, \hat{v_j}, o_1, o_2) = \left\{ 
						\begin{array}{lr}
						& \text{$x=\hat{x}, y=\hat{y}$} \\
						(\hat{x},\hat{y}, \rhd, \rhd) 
							 & \text{IF $\hat{x} \leq \hat{y},$} \\
					    (\hat{y},\hat{x}, \lhd, \lhd) & \text{ELSE} \\
						& \\

						& \text{$x\neq\hat{x} \wedge y=\hat{y}$} \\
						(\hat{x},\hat{y}, \lhd, \rhd) & \text{IF $\hat{x} \leq \hat{y}$} \\ 
						(\hat{y},\hat{x}, \lhd, \rhd) & \text{ELSE} \\ 

						& \\
						& \text{$x=\hat{x} \wedge y\neq\hat{y}$} \\
						(\hat{x},\hat{y}, \rhd, \lhd) & 
							\text{IF $\hat{x} \leq \hat{y}$} \\ 
						(\hat{y},\hat{x}, \rhd, \lhd) & 
							\text{ELSE} \\ 

						& \\
						& \text{$x\neq\hat{x} \wedge y\neq\hat{y}$} \\
						(\hat{x},\hat{y}, \lhd, \lhd) & 
							\text{IF $\hat{x} \leq \hat{y}, $} \\ 
						(\hat{y},\hat{x}, \rhd, \rhd) & 
							\text{ELSE} \\ 
						\end{array}
						\right.
\end{array}
$$

\item {\sf [STEP-2] }{\bf Reduce multiplicity:} Sort all the bi-directed edges 
in {\sf [STEP-1]}, using radix sort. Since the parameter $k$ is always odd this
guarantees that a pair of canonical $k$-mers have exactly one orientation. Remove
the duplicates and record the multiplicities of each canonical edge. Gather all the
unique canonical edges into an edge list $\mathcal{E}$.

\item {\sf [STEP-3] }{\bf Collect bi-directed vertices:} For each canonical bi-directed edge
$(\hat{v_i},\hat{v_j},o_1, o_2) \in \mathcal{E}$, collect the canonical $k$-mers
$\hat{v_i}$, $\hat{v_j}$ into list $\mathcal{V}$. Sort the list $\mathcal{V}$ and
remove duplicates, such that $\mathcal{V}$ contains only the unique canonical
$k$-mers.

\item {\sf [STEP-4] }{\bf Adjacency list representation:} 
The list $\mathcal{E}$ is the collection of all the edges in the bi-directed graph 
and the list $\mathcal{V}$ is the collection of all the nodes in the bi-directed 
graph. It is now easy  to use $\mathcal{E}$ and generate the adjacency 
lists representation for the bi-directed graph. This may require one extra
radix sorting step.
\end{itemize}
\section{Analysis of the algorithm {\sf BiConstruct}}
\begin{theorem}
Algorithm {\sf BiConstruct} builds a bi-directed de Bruijn graph of the order $k$ in
$\Theta(n)$ time. Here $n$ is number of characters/symbols in the input.
\end{theorem}
\begin{proof}
Without loss of generality assume that all the reads are of the same size $r$. Let $N$ be
the number of reads in the input. This generates a total of $(r-k)N$ $(k+1)$-mers in {\sf [STEP-1]}.
The radix sort needs to be applied at most $2k\log(|\Sigma|)$ passes, resulting in $2k\log(|\Sigma|)(r-k)N$
operations. Since $n=Nr$ is the total number of characters/symbols in the input, the radix sort
takes $\Theta(kn\log(|\Sigma|))$ operations assuming that in each pass of sorting only a constant number of symbols are used. If $k\log (|\Sigma|)=O(\log N)$, the sorting takes only $O(n)$ time. In practice since $N$ is very large in relation to $k$ and $|\Sigma|$, the above condition readily holds. Since the time for this step dominates that of all the other steps,
the runtime of the algorithm {\sf BiConstruct} is $\Theta(n)$.
\end{proof}

\section{Parallel algorithm for building bi-directed de Bruijn graph}
\label{sec:parallel}
In this section we illustrate a parallel implementation of our algorithm.
Let $p$ be the number of processors available. We first distribute $N/p$ reads for each processor. All 
the processors can execute {\sf [STEP-1]} in parallel. In {\sf [STEP-2]} we need to perform 
parallel sorting on the list $\mathcal{E}$. Parallel radix/bucket sort --which does not use 
any all-to-all communications-- can be employed to accomplish this. For example, the integer sorting algorithm of
Kruskal, Rudolph and Snir takes $O\left (\frac{n}{p}\frac{\log n}{\log(n/p)}\right )$ time. This will be $O(n/p)$ if $n$ is
a constant degree polynomial in $p$. In other words, for coarse-grain parallelism the run time is asymptotically
optimal. In practice coarse-grain parallelism is what we have. Here $n = N(r-k+1)$.
We call this algorithm {\sf Par-BiConstruct}. 

\begin{theorem}
Algorithm {\sf Par-BiConstruct} builds a bi-directed de Bruijn graph in time $O(n/p)$. The message complexity is
$O(n)$.
\end{theorem}

\subsection{Some remarks on Jackson and Aluru's algorithm}
\label{sec:remarks}
The algorithm of Jackson and Aluru~\cite{par_bidirected_graph}  first identifies the vertices of the bi-directed
graph -- which they call representative nodes. Then for every representative node $|\Sigma|$ 
many-to-many messages were generated. These messages correspond to potential bi-directed edges
which can be adjacent on that representative node. A bi-directed edge is successfully created
if both the representatives of the generated message exist in some processor, otherwise the
edge is dropped. This results in generating a total of $\Theta(n|\Sigma|)$ many-to-many messages.
The authors in the same paper demonstrate that communicating many-to-many messages is a major 
bottleneck and does not scale well. On the other had we remark that the algorithm {\sf BiConstruct}
does not involve any many-to-many communications and does not have any scaling bottlenecks. 

On the other hand the algorithm presented in their paper~\cite{par_bidirected_graph} can potentially 
generate spurious bi-directed edges. According to the definition~\cite{bidirected_graph} of the 
bi-directed de Bruijn graph in the context of SA problem, a bi-directed edge between two $k$-mers/vertices
exists iff there exists some read in which these two $k$-mers are adjacent. We illustrate this by a simple
example. Consider a read $r_i = AATGCATC$. If we wish to build a bi-directed graph of order $3$, then
$\{AAT, ATG, TGC, GCA, CAT, ATC\}$ form a subset of the vertices of the bi-directed graph. In this example
we see that $k$-mers $AAT$ and $ATC$ overlap by exactly $2$ symbols. However there cannot be any 
bi-directed edge between them according to the definition -- because they are not adjacent. On the other hand
the algorithm presented in ~\cite{par_bidirected_graph} generates the following edges with respect
to $k$-mer $AAT$: $\{(AAT,ATA), (AAT,ATG), (AAT,ATT)$, $(AAT,ATC) \}$. The edges $(AAT, ATA)$ and $(AAT, ATC)$
are purged since the $k$-mers $ATA$ and $ATC$ are missing. However bi-directed edges with corresponding orientations
are established between $ATG$ and $ATC$. Unfortunately $(AAT, ATC)$ is a spurious edge and can potentially
generate wrong assembly solutions. In contrast to their algorithm~\cite{par_bidirected_graph} our algorithm 
does not use all-to-all communications -- although we use point-point communications.

\section{Out of core algorithms for building bi-directed de Bruijn graphs} 
\label{sec:out-of-core}
\begin{theorem}
There exists an out-of core algorithm to construct a bi-directed de Bruijn graph using an optimal number of
I/O's.
\end{theorem}
\begin{proof}
{\em Sketch:} Replace the radix sorting with an external $R-$way merge which takes only 
$\Theta(\frac{n\log(n/B)}{B\log(M/B)})$. Where $M$ is the main memory size, $n$ is the sum 
of the lengths of all reads, and $B$ is the block size of the disk.
\end{proof}

\begin{figure}
\begin{center}
\includegraphics[scale=0.5]{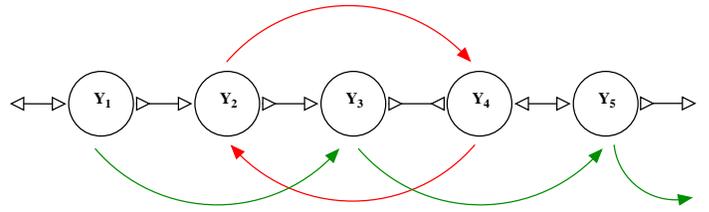}
\end{center}
\caption{Problems with pointer jumping on bi-directed chains}
\label{fig:bi-pointer}
\end{figure}

\section{Simplified bi-directed de Bruijn graph}
\label{sec:simplification}
\begin{figure}
\begin{center}
\includegraphics[scale=0.6]{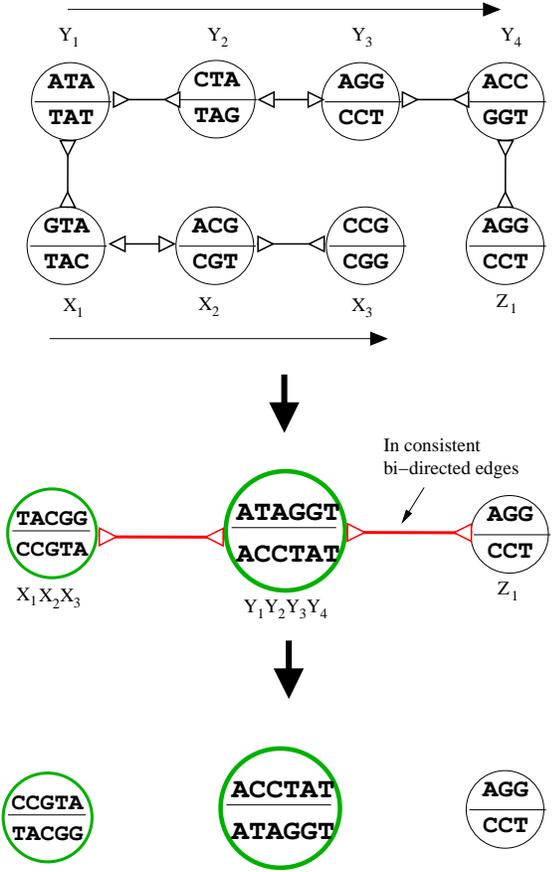}
\end{center}
\caption{Issues with partially compacted bi-directed chains}
\label{fig:chains}
\end{figure}

The bi-directed de Bruijn graph constructed in the previous section may contain several linear 
chains. These chains have to be compacted to save space as well as time. The graph that results
after this compaction step is refered to as
the {\em simplified bi-directed graph}. A linear chain of bi-directed edges between nodes 
$u$ and $v$ can be compacted only if we can find a valid bi-directed walk connecting $u$ and $v$. 
All the $k$-mers/vertices in a compactable chain can be 
merged into a single node, and relabelled with the corresponding forward and reverse complementary strings. 
In Figure~\ref{fig:chains} we can see that the nodes $X_1$ and $X_3$ can be connected with a valid bi-directed 
walk and hence these nodes are merged into a single node. In practice the compaction of chains seems
to play a very crucial role. It has been reported that merging the linear chains can 
reduce the number of nodes in the graph by up-to $30\%$~\cite{velvet08}. 

\begin{figure}
\begin{center}
\scalebox{0.6}{\input{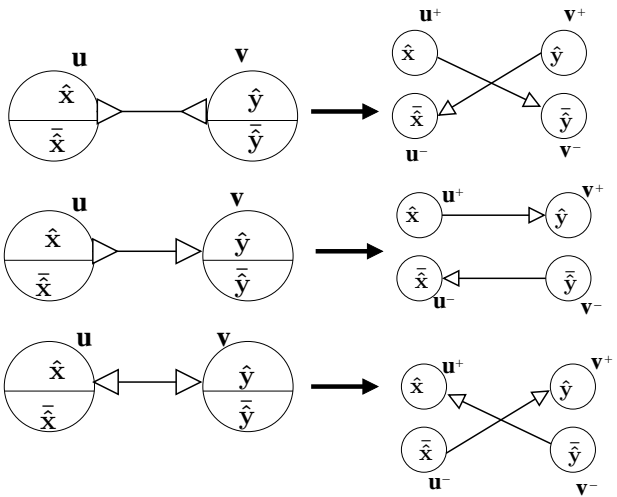}}
\end{center}
\caption{Transforming bi-directed chain compaction to list ranking}
\label{fig:chain-compact}
\end{figure}

Although bi-directed chain compaction problem seems like a {\em list ranking} problem
there are some fundamental differences. Firstly, a bi-directed edge can be traversed in
both the directions. As a result, applying {\em pointer jumping} directly on a bi-directed
graph can lead to cycles and cannot compact the bi-directed chains correctly. 
Figure~\ref{fig:bi-pointer} illustrates the first phase of pointer jumping. As we can
see, the {\em green} arcs indicate valid pointer jumps from the starting nodes. However since
the orientation of the node $Y_4$ is reverse relative to the direction of pointer jumping
a cycle results. In contrast, a valid bi-directed chain compaction would merge
all the nodes between $Y_1$ and $Y_5$ since there is a valid bi-directed walk between $Y_1$
and $Y_5$. On the other hand, bi-directed chain compaction may result in inconsistent 
bi-directed edges and these edges should be recognised and removed. Consider a bi-directed
chain in Figure~\ref{fig:chains}; this chain contains two possible bi-directed walks -- $Y_1$ to 
$Y_4$ and $X_1$ to $X_3$. The walk from $Y_1$ to $Y_4$ ($Y_4$ to $Y_1$) spells out a 
label $ATAGGT$($ACCTAT$) after compaction. Once we perform this compaction the edge between $Y_4$
and $Z_1$ in the original graph is no longer valid -- because the $k$-mer on $Z_1$ cannot overlap
with the label $ACCTAT$. The same is true for the compaction of the bi-directed walk between $X_1$
and $X_3$. The redundant edges after compaction are marked in {\em red}. Since bi-directed chain
compaction has a lot of practical importance efficient and correct algorithms are essential. 

\begin{figure}
\begin{center}
\includegraphics[scale=0.6]{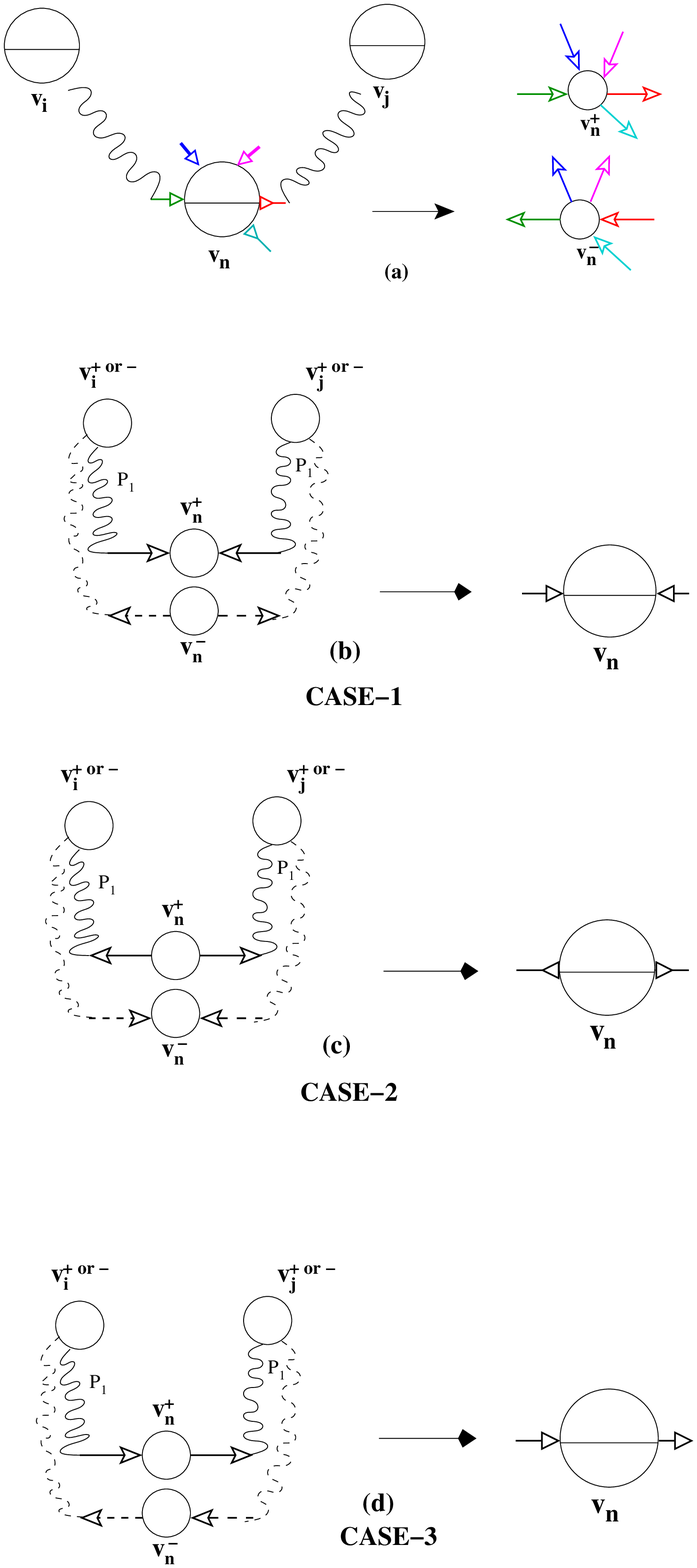}
\end{center}
\caption{Proof that {\sf ListRankingTransform} preserves bi-directed walk in the original graph.}
\label{fig:proof}
\end{figure}

We now provide algorithms for the bi-directed chain compaction problem. Our key idea here is
to transform a bi-directed graph into a directed graph and then apply {\em list ranking}.
Given a list of candidate canonical bi-directed edges, we apply a {\sf ListRankingTransform}
(see Figure~\ref{fig:chain-compact}) which introduces two new nodes $v^+,v^-$ for every node
$v$ in the original graph. Directed edges corresponding to the orientation are introduced. See 
Figure~\ref{fig:chain-compact}. 
\begin{lemma}
\label{lem:bi-lemma}
Let $BG(V,E)$ be a bi-directed graph; let $BG^t(V^t,E^t)$ be the directed graph after applying
{\sf ListRankingTransform}. Two nodes $u,v \in V$ are connected by a bi-directed path iff 
$u^+\in V^t$ ($u^-\in V^t$) is connected to one of $v^+$($v^-$) or $v^-$($v^+$) by a directed path. 
\end{lemma}
\begin{proof}
We first prove the forward direction by induction on the number of nodes in the bi-directed graph. 
Consider the {\em basis} of induction when $|V|=2$, let $v_0,v_1 \in V$. Clearly we are only interested
when $v_0$ and $v_1$ are connected by a bi-directed edge. By the definition of {\sf ListRankingTransform} 
the Lemma in this case is trivially true. Now consider a bi-directed graph with $|V|=n+1$ nodes, if the
path between $v_i, i<n$ and $v_j, j<n$ does not involve node $v_n$ the lemma still holds by induction on 
the sub bi-directed graph $BG(V-\{v_n\},E)$. Now assume that $v_i\ldots v_p,v_n,v_q\ldots v_j$ is the 
bi-directed path between $v_i$ and $v_j$ involving the node $v_n$; see Figure~\ref{fig:proof}(a). 
Also Figure~\ref{fig:proof}(a) shows how the transformed directed graph look like; observe the colors of 
bi-directed edges and corresponding directed edges. By induction hypothesis on the sub bi-directed paths
$v_i\ldots v_p,v_n$ and $v_n,v_q\ldots v_j$ we have the following. $v_i^+$ is connected to $v_n^+$
or $v_n^-$ by some directed path $P_1$ (See Figure~\ref{fig:proof}(b); $v_n^+$ is connected to $v_j^+$ 
or $v_j^-$ by some directed path $P_2$. We examine three possible cases depending on how the directed edge 
from $P_1$ and $P_2$ is incident on $v_n^+$. In {\sf CASE-1} we have both $P_1$ and $P_2$ pointing into
node $v_n^+$. This implies that the orientation of the bi-directed edges in the original graph is
according to Figure~\ref{fig:proof}(b). In this case we cannot have a bi-directed walk involving the node
$v_n$, which contradicts our original assumption. Similarly {\sf CASE-2}(Figure~\ref{fig:proof}(c)) would 
also lead to a similar contradiction. Only {\sf CASE-3} would let node $v_n$ involve in a bi-directed walk.
In this case $v_i^+$ will be connected to either $v_j^+$ or $v_j^-$ by concatenation of the paths $P_1,P_2$.
 We can make a similar argument to prove the reverse direction.
\end{proof}
\subsection{Algorithm for bi-directed chain compaction}
We first identify a set of candidate bi-directed edges which can potentially form a chain. One possible
criterion will be to include all the edges which are adjacent on bi-directed nodes with exactly one in
and out degree. Each candidate bi-directed edge is transformed using {\sf ListRankingTranform} and 
{\sf list ranking} is applied on resultant set. As a consequence of the symmetry in {\sf ListRankingTransform}
we would see both forward and reverse complements of the compacted chains in the output. We can further
canonicalize each chain and remove the duplicates by sorting. This results in unique bi-directed chains
from the candidate bi-directed edges. Finally we report only the chains which are resultant of compaction
of at least three bi-directed nodes. This removes all the inconsistent edges (see Figure~\ref{fig:chains})
from further consideration. As a consequence of Lemma~\ref{lem:bi-lemma} all the bi-directed chains are
correctly compacted.
\subsection{Analysis of bi-directed compaction on parallel and out-of-core models}
Let $\mathcal{E}_l$ be the list of candidate edges for compaction. To do compaction in parallel, we can use 
a {\em Segmented Parallel Prefix} on $p$ processors to accomplish this in time $O(|2\mathcal{E}_l|/p +\log(p))$. 
On the other hand list ranking can also be done out-of-core as follows. Without loss of generality we can
treat the input for the list ranking problem as a set $S$ of ordered tuples of the form $(x,y)$. Given $S$
we create a copy and call it $S'$. We now perform an external sort of $S$, $S'$ with respect to $y$ (i.e., using
the $y$ value of tuple $(x,y)$ as the key) and $x$ respectively. The two sorted lists are scanned linearly to
identify tuples $(x, y)\in S$, $(x', y')\in S'$ such that $y=x'$. These two tuples are merged into a single
tuple $(x,y')$ and are added to a list $\mathcal{E}'_l$. This process is  now repeated on $\mathcal{E}'_l$.
Note that if the underlying graph induced by $\mathcal{E}_l$ does not have any cycles then 
$|\mathcal{E}'_l| \leq |\mathcal{E}_l|/2$; which means that the size of $\mathcal{E}'_l$ geometrically
decreases after every iteration. The I/O complexity of each iteration is dominated by the external sorting and hence bi-directed compaction can be accomplished out-of-core with $\Theta(|\mathcal{E}_l|/B\log_{\frac{M}{B}}(|\mathcal{E}_l|/B))$ I/O operations.

Care should be taken to deal with cycles. There are two ways of dealing with cycles. One way is to first identify
all the cycles in the bi-directed graph (generated in the previous section) and then break these cycles by removing
appropriate edges. A second easy is to identify the cycles on the fly and break them. We employ the second
approach. 

\section{Improving the construction of the bi-directed de Bruijn graph in some 
practical assemblers}
\label{sec:velvet}
In this section we briefly describe how our algorithms can be used to speedup some of the
existing SA programs. As an example, we consider {VELVET}~\cite{velvet08}. VELVET is a suite of 
programs -- {\sf velveth} and {\sf velvetg}, which has recently gained acclamation in assembling 
short reads. VELVET program builds a simplified bi-directed graph from a set of reads. We now
briefly describe the algorithm used in VELVET to build this graph. VELVET program puts all the 
$k$-mers from the input into a hash table and then identifies the $k$-mers which are present 
in at least 2 reads -- this information is called the {\em roadmap} in VELVET's terminology.  
The program then builds a de Bruijn graph using these $k$-mers. Finally it takes every read 
and threads it on these $k$-mers. The worst case time complexity is $O(n\log(n))$ -- assuming 
that $k$ and $|\Sigma|$ are constants. On the other hand since VELVET
builds this graph entirely in-memory, this has some serious scalability problems especially on 
large scale assembly projects. However VELVET seems to have some very good assembly 
heuristics to remove errors and identify redundant assembly paths, etc. Thus our algorithm can 
act as a replacement to code in VELVET which performs in-memory graph construction. 


\section{Experimental results}
\label{sec:exp_results}
We have compared the performance of our algorithm and that of Jackson and Aluru~\cite{par_bidirected_graph}.
We refer to the later algorithm as JA.
To make this comparison fair, we have implemented the JA algorithm also on the same platform that
our algorithm runs on.
We have used a SGI/Altix IA-64 machine with 64 nodes for all of our experiments. 
Our implementation uses MPI for communication between the processors. Table~\ref{tab:aluru_comparison} 
shows the user and system times for both our algorithm and the JA algorithm. We can clearly see 
that the system time (communication time) is consistently higher for the JA algorithm. Also, as we move from 
one million to eight million reads the increase in the system time is quite significant for the JA
algorithm (e.g., the system time for JA increases from 0.621 sec to 4.120 sec, which
is almost a 7X increase. On the other hand there is only a 3X increase in our algorithm). The JA
algorithm has a higher communication cost because it enumerates all the bi-directed edges and uses 
many-to-many messages to send to the right processor. 

The user time of our algorithm is
also consistently superior compared to the user time of JA. This clearly is because
we do much less local computations. In contrast, JA needs to do a lot of local processing
which arises from processing all the received edges, removing redundant ones, and collecting the necessary
edges to perform many-to-many communications. Since JA was taking a significant amount of
time on for inputs larger than 8 million we have compared these algorithms only for input sizes
up to 8 million. The experimental
results reported in \cite{par_bidirected_graph} start with at least 64 processors. We however
show the scalablity of our algorithm upto $128$ million reads in Table~\ref{tab:scalability}.
Table~\ref{tab:scalability} clearly demonstrates the scalability of our algorithm. We make our implementations available
at \url{http://trinity.engr.uconn.edu/~vamsik/ParBidirected.tgz}.

\begin{table}
\begin{center}
\begin{tabular}{|c|c|c|c|c|c|}
\hline
\text{p}&\multicolumn{2}{|c|}{JA ALGO}&\multicolumn{2}{|c|}{OUR ALGO} &\\
\hline
&\text{user time}&\text{sys time}&\text{user time}&\text{sys time}&\text{speed up} \\ 
&\text{(sec)}&\text{(sec)}&\text{(sec)}&\text{(sec)}  & \\  
\hline
\multicolumn{6}{|c|}{\text{READS=$1048576$}} \\ 
\hline
4 & 55.932 & 0.621 & 2.365 & 0.046 & 23.456\\ 
\hline
8 & 25.161 & 0.331 & 3.072 & 0.035 & 8.205\\ 
\hline
16 & 13.603 & 0.175 & 0.619 & 0.038 & 20.971\\ 
\hline
32 & 5.711 & 0.157 & 0.149 & 0.099 & 23.661\\ 
\hline
\multicolumn{6}{|c|}{\text{READS=$8388608$}} \\ 
\hline
4 & 593.712 & 4.120 & 20.807 & 0.159 & 28.514\\ 
\hline
8 & 341.694 & 2.322 & 17.637 & 0.105 & 19.390\\ 
\hline
16 & 147.629 & 1.117 & 17.734 & 0.087 & 8.347\\ 
\hline
32 & 72.413 & 0.566 & 13.967 & 0.120 & 5.181\\ 
\hline
\end{tabular}

\end{center}
\caption{Comparision between the JA algorithm and our algorithm}
\label{tab:aluru_comparison}
\end{table}

\begin{table}
\begin{center}
\begin{tabular}{|c|c|c|c|}
\hline
\text{p} & \text{user time} & \text{sys time} & \text{wall time}\\
 & \text{(ticks)} & \text{(ticks)} & \text{(min:sec)}\\
\hline
\hline
\multicolumn{4}{|c|}{\text{READS=$16777216$}}\\
\hline
2 & 37147 & 259 & 1:14.02 \\
\hline
4 & 37254 & 85 & 0:38.95 \\
\hline
8 & 20217 & 57 & 0:21.90 \\
\hline
16 & 16951 & 55 & 0:19.73 \\
\hline
32 & 12901 & 40 & 0:16.38 \\
\hline
\hline
\multicolumn{4}{|c|}{\text{READS=$33554432$}}\\
\hline
2 & 148070 & 1219 & 2:42.66 \\
\hline
4 & 99067 & 677 & 1:48.60 \\
\hline
8 & 47319 & 322 & 0:55.41 \\
\hline
16 & 17936 & 135 & 0:25.64 \\
\hline
32 & 9973 & 191 & 0:17.55 \\
\hline
\hline
\multicolumn{4}{|c|}{\text{READS=$67108864$}}\\
\hline
2 & 340653 & 2348 & 6:18.77 \\
\hline
4 & 240861 & 1931 & 4:14.57 \\
\hline
8 & 153782 & 1781 & 2:39.18 \\
\hline
16 & 64408 & 804 & 1:10.91 \\
\hline
32 & 46659 & 486 & 0:53.32 \\
\hline
\hline
\multicolumn{4}{|c|}{\text{READS=$134217728$}}\\
\hline
2 & 770922 & 5560 & 15:00.42 \\
\hline
4 & 471196 & 4272 & 8:29.62 \\
\hline
8 & 314281 & 3456 & 5:17.65 \\
\hline
16 & 135562 & 2148 & 2:21.83 \\
\hline
32 & 82414 & 950 & 1:28.87 \\
\hline
\end{tabular}

\end{center}
\caption{Scalablility of our algorithm for up to 128 million reads}
\label{tab:scalability}
\end{table}

\section{Conclusions}
In this paper we have presented an efficient algorithm to build a bi-directed de Bruijn graph which is a fundamental
data structure for any sequence assembly program -- based on Eulerian approach. Our algorithms are also efficient
 in parallel and out of core settings. These algorithms can be used in building large scale bi-directed de Bruijn graphs. 
Also, our algorithm does not employ any all-to-all communications in parallel setting and performs better than that
of Jackson and Aluru~\cite{par_bidirected_graph}.

\noindent {\bf Acknowledgements.} This work has been supported in
part by the following grants: NSF 0326155, NSF 0829916 and NIH
1R01GM079689-01A1.
\bibliographystyle{IEEEtran}
\bibliography{vamsi_algos.bib}

\end{document}